\documentclass[reprint,english,aps,showpacs]{revtex4-1}

\pdfoutput=1
\usepackage{amsmath}
\usepackage{color}
\usepackage{graphicx}
\usepackage{epstopdf}

\usepackage{babel}

\begin{document}

\title{Long-range order and thermal stability of thin Co$_{2}$FeSi films on GaAs(111)B}

\author{B.~Jenichen}
\email{bernd.jenichen@pdi-berlin.de}

\author{J.~Herfort}

\author{K.~Kumakura}
\altaffiliation[Also at ]{NTT Basic Research Laboratories, 3-1
Morinosato Wakamiya, Atsugi-shi, Kanagawa 243-0198, Japan.}

\author{A.~Trampert}

\affiliation{Paul-Drude-Institut fuer Festkoerperelektronik,
Hausvogteiplatz 5--7,D-10117 Berlin, Germany}

\date{\today}
\begin{abstract}
Co$_{2}$FeSi/GaAs(111)B hybrid structures are grown by
molecular-beam epitaxy and characterized by transmission electron
microscopy (TEM) and x-ray diffraction. The Co$_{2}$FeSi films
grow in an island growth mode at substrate temperatures $T_{S}$
between $T_{S}$~=~100$\thinspace^{\circ}$C and
425$\thinspace^{\circ}$C. The structures have a stable interface
up to $T_{S}=275~^{\circ}$C. The films contain
 fully ordered $L2_{1}$ and partially ordered $B2$ phases.
The spatial distribution of long-range order in Co$_{2}$FeSi is
characterized using a comparison of TEM images taken with
superlattice reflections and the corresponding fundamental
reflections. The spatial inhomogeneities of long-range order can
be explained by local non-stoichiometry due to lateral segregation
or stress relaxation without formation of extended defects.
\end{abstract}


\maketitle

\section{Introduction}
There is an increasing interest in Heusler alloys as candidates
for sources of spin injection into semiconductors
\cite{Felser09,prinz98,fert08,zutic04}. The Heusler alloy
Co$_{2}$FeSi is a ferromagnetic half-metal with a Curie
temperature larger than 1100~K \cite{wurmehl05,wurmehl06}. The
lattice parameter matches that of GaAs. Therefore, Co$_{2}$FeSi is
a suitable material for spin injection into GaAs-based structures
such as for example spin light-emitting diodes (spin-LEDs
\cite{Ohno99,zhu01,Hanbicki02,ramsteiner08}) and spin field effect
transistors \cite{Sugahara04}. Samples produced by radio frequency
magnetron sputtering have shown magnetically dead layers near the
interfaces \cite{Kallmayer06}. An epitaxial growth technique like
molecular-beam epitaxy (MBE) may overcome these difficulties.

Recent theoretical studies indicate that a ${\{111\}}$~interface
may be superior to ${\{001\}}$ interfaces \cite{Attema06} with
respect to the half-metallic properties of Co$_{2}$FeSi. Moreover,
reacted compounds sometimes developing during Co$_{2}$FeSi growth
on GaAs(001) are bounded by ${\{111\}}$ planes \cite{hashimoto07}.
This observation led to the hypothesis that a ${\{111\}}$
interface would be more stable. At the same time, reduced
diffusion is expected perpendicular to the interface during growth
on GaAs(111)B.

X-ray and electron diffraction experiments yield information about
structure and long-range order of Heusler alloys and related
materials
\cite{niculescu76,niculescu79,Jenichen05,wurmehl06,takamura09}. In
this study MBE grown Co$_{2}$FeSi/GaAs(111)B hybrid structures are
investigated by transmission electron microscopy (TEM), x-ray
diffraction and atomic force microscopy (AFM) in order to
characterize the structural properties and the stability of the
ferromagnet/semiconductor (FM/SC) interface and the Co$_{2}$FeSi
film. The growth mode of the epitaxial film
\cite{bauer58,Tsao1993,kag09,Jenichen09} is expected to have a
fundamental influence on its structural properties.


\begin{figure}[b]
 \includegraphics[width= 8cm]{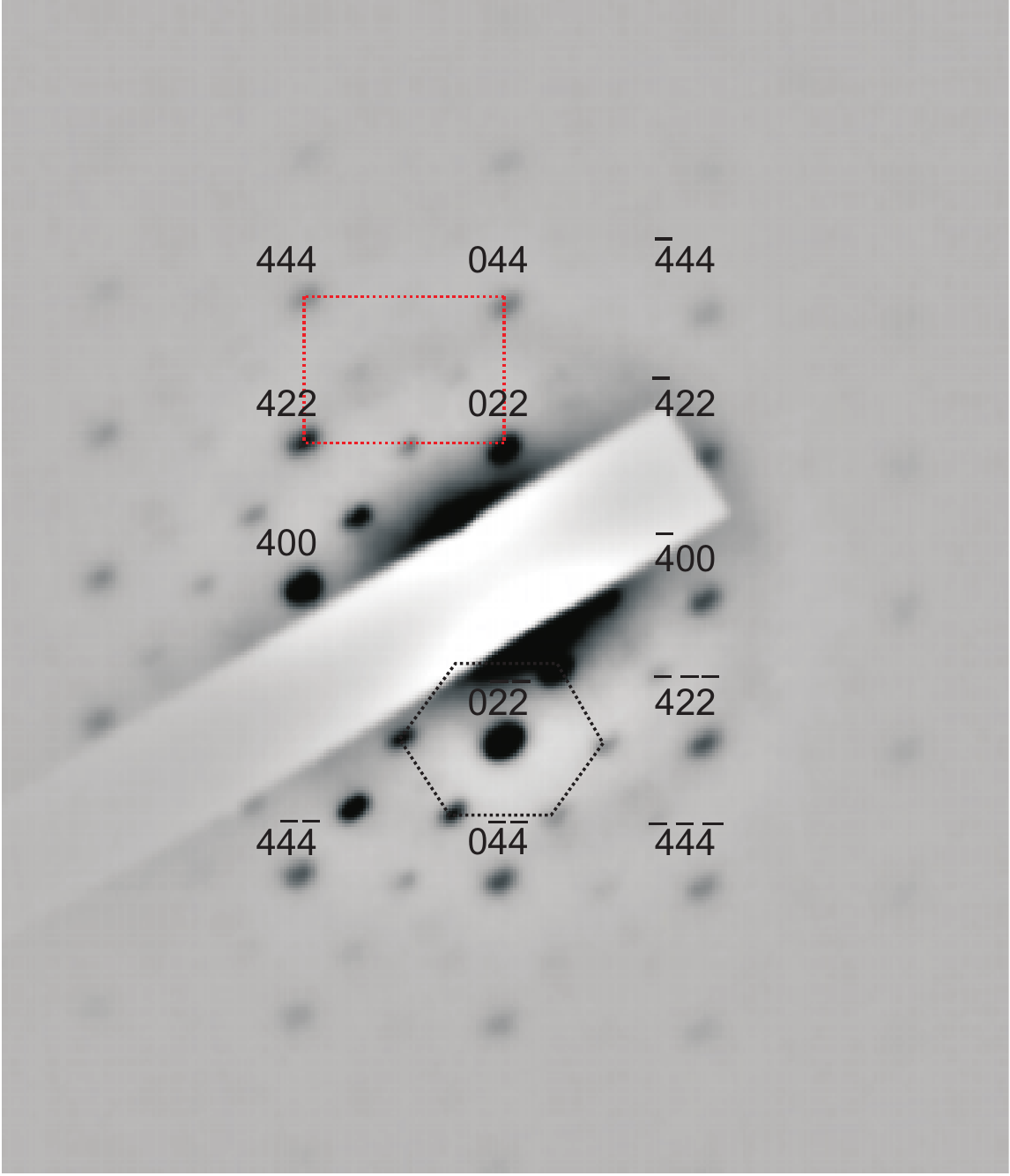}

\caption{Selected area diffraction pattern along the
Co$_{2}$FeSi/GaAs~{[}0$1\bar{1}${]} zone axis of a sample grown at
$T_{S}$~=~275$\thinspace^{\circ}$C. A rectangular pattern of
Co$_{2}$FeSi fundamental reflections (indexed, red dashed line) of
the $L2_{1}$ structure is observed. This rectangular pattern is
slightly more pronounced than the hexagonal pattern (black dashed
line) of the superlattice maxima in the vicinity. }

\label{fig:figure1}
\end{figure}

%
\begin{figure}[b]
 \includegraphics[width= 8cm]{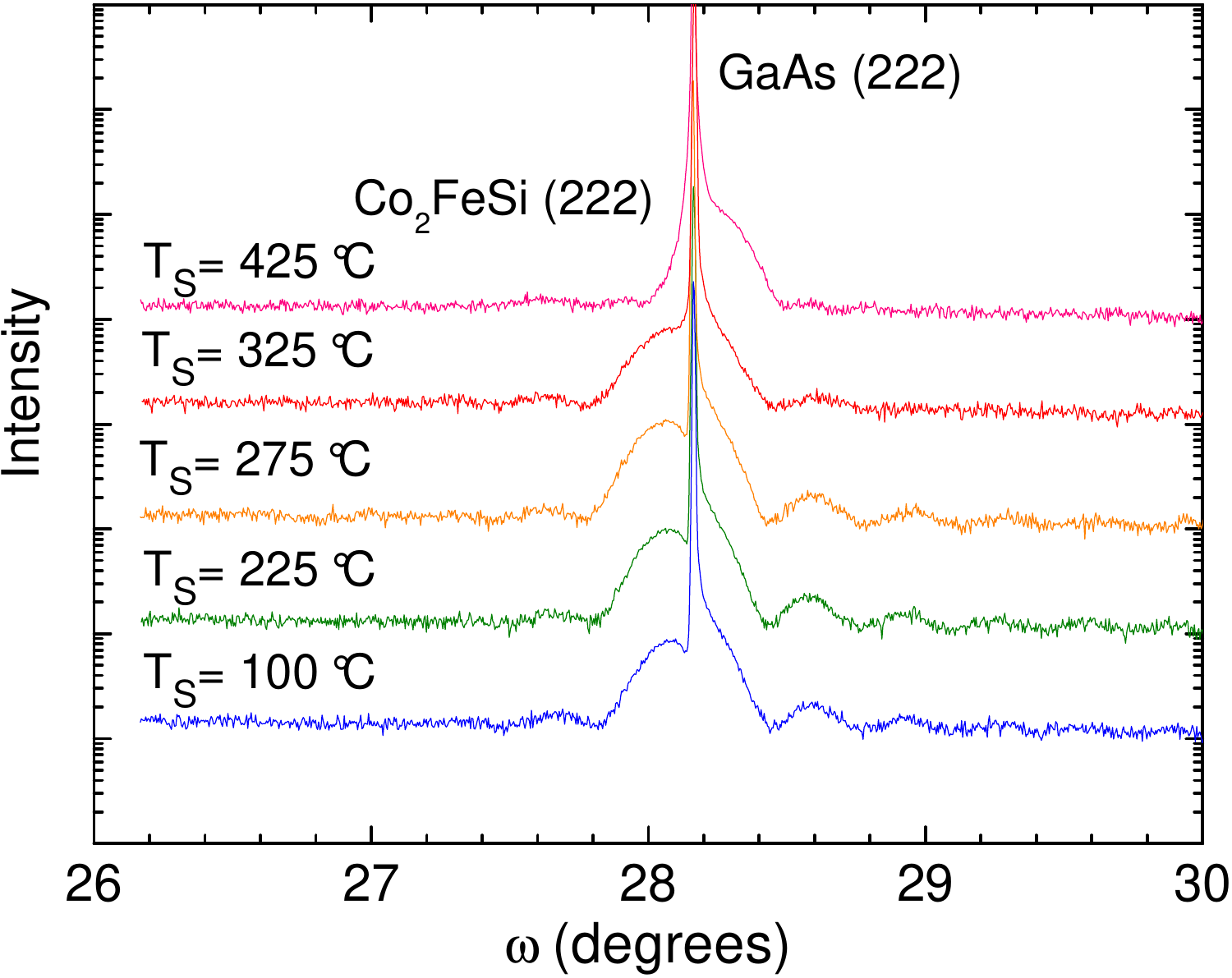}

\caption{High resolution x-ray diffraction curves measured near
the GaAs 222 reflection of samples grown at different substrate
temperatures $T_S$. The narrow peak is the quasiforbidden GaAs
reflection and the broader maximum is the Co$_{2}$FeSi 222
superlattice reflection indicating at least a $B2$ ordering of all
the films. The thickness fringes demonstrate a high quality of the
FM/SC interface and surface.}

\label{fig:figure2}
\end{figure}

%
\begin{figure}[b]
 \includegraphics[width = 8cm]{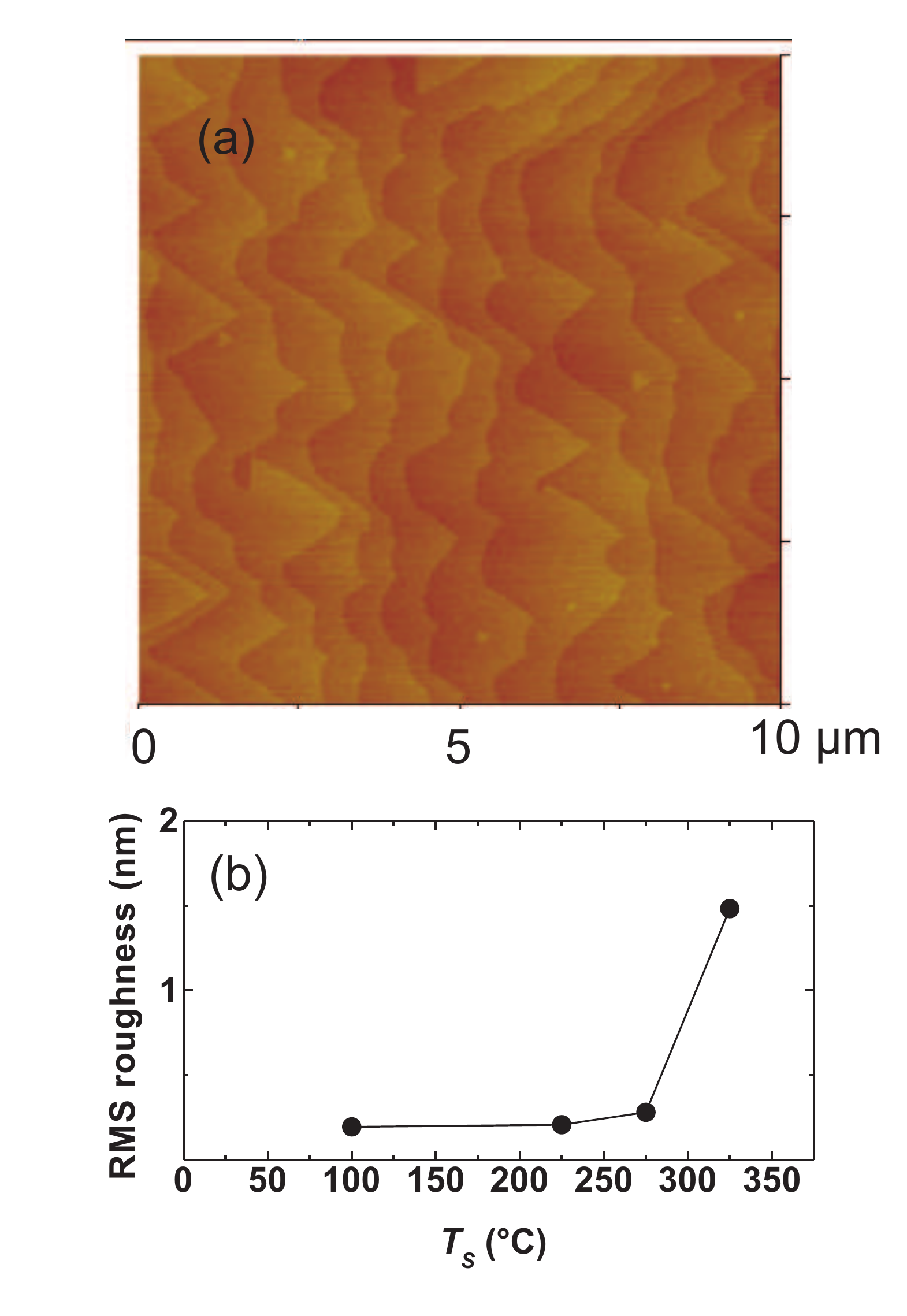}

\caption{(a) AFM micrograph of the Co$_{2}$FeSi surface grown on
GaAs(111)B at $T_S$= 100$\thinspace^{\circ}$C. The pattern of the
steps is similar to that of a GaAs(111)B surface. Some triangular
islands are visible on the broad terraces. (b) Root mean squared
(RMS) surface roughness in dependence of the growth temperature of
the samples. }

\label{fig:figure3}
\end{figure}

%
\begin{figure}[b]
\includegraphics[clip,width= 8cm]{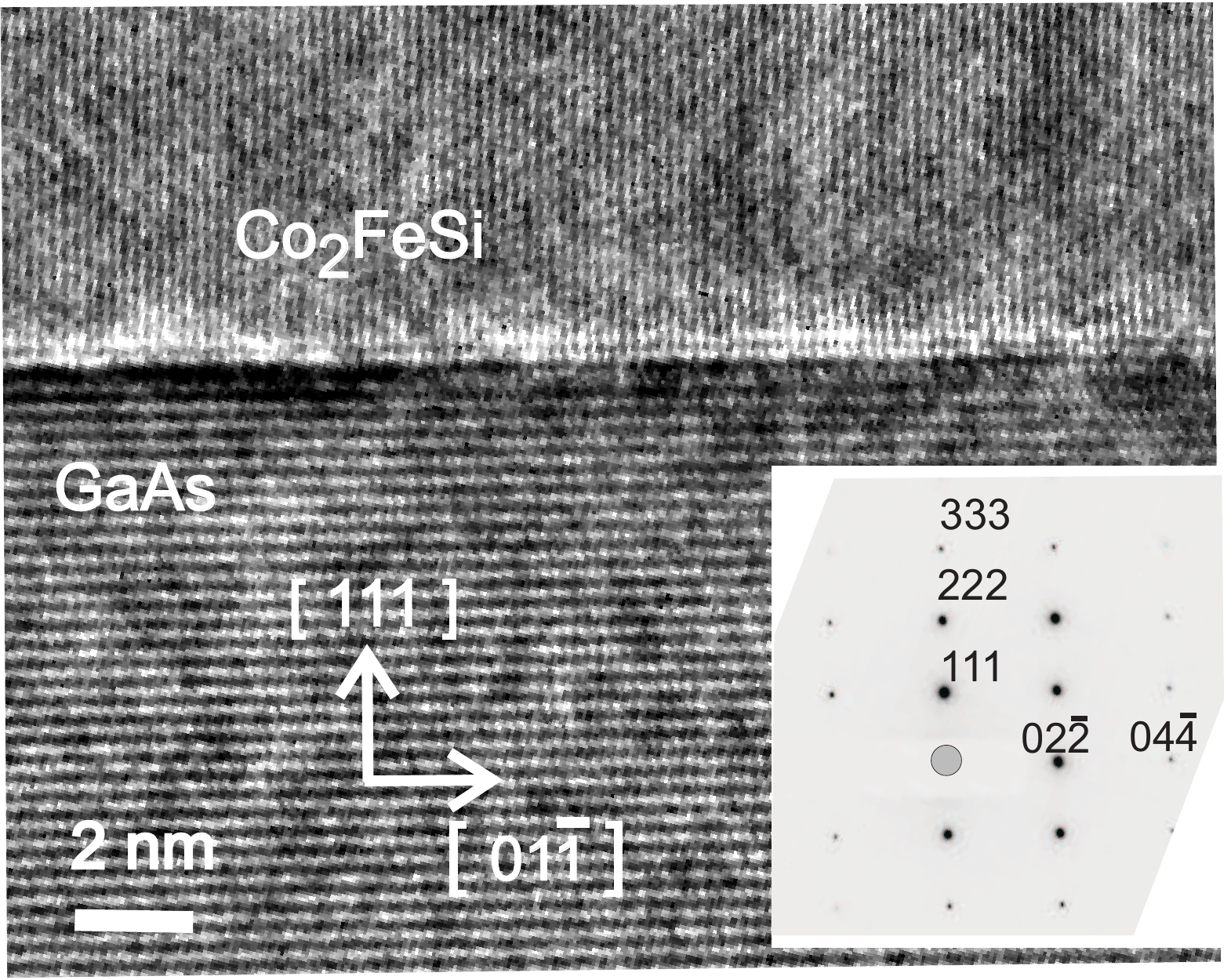}\\

\caption{ Cross-section HRTEM micrograph (Fourier filtered) along
the GaAs {[}2$\bar{1}\bar{1}${]} zone axis of a sample grown at
$T_{S}$~=~100$\thinspace^{\circ}$C and a corresponding selected
area diffraction pattern illustrating the orientational
relationship between Co$_{2}$FeSi and GaAs. Due to the vanishing
misfit between stoichiometric Co$_{2}$FeSi and GaAs, their
diffraction maxima fully coincide. The interface is smooth. }

\label{fig:figure4}
\end{figure}

%
\begin{figure}[b]
\includegraphics[clip,width= 8cm]{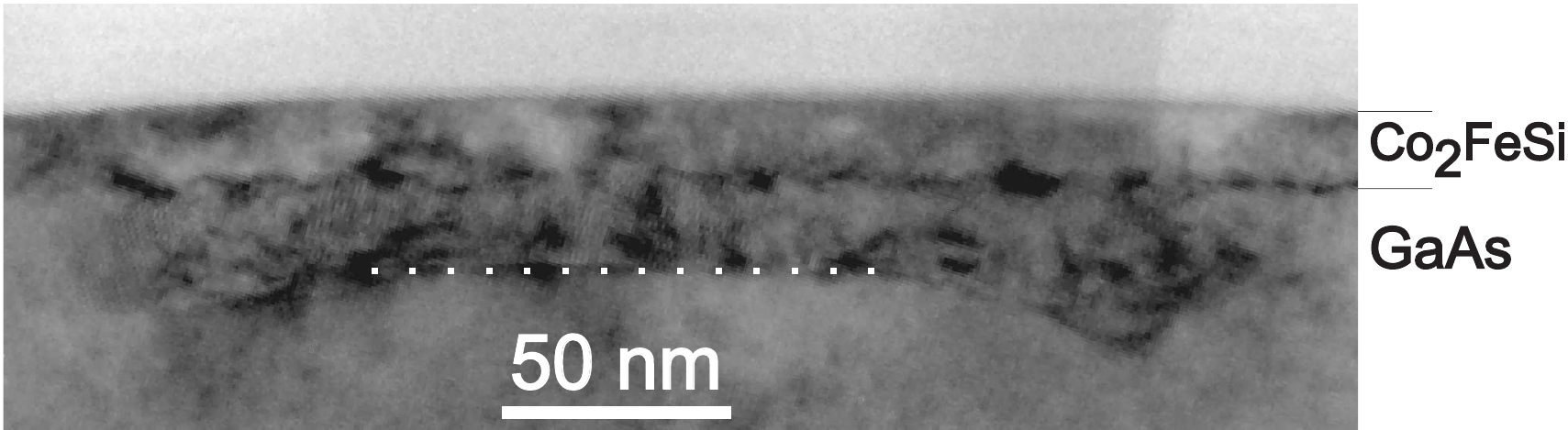}\\

\caption{Cross-section multi-beam TEM micrograph along the GaAs
{[}0$1\bar{1}${]} zone axis of a Co$_{2}$FeSi film grown on GaAs
at $T_{S}$~=~325$\thinspace^{\circ}$C. A large reacted region
(lower edge marked by dotted line) is visible below the
Co$_{2}$FeSi/GaAs(111) interface . }

\label{fig:figure5}
\end{figure}

%
\begin{figure}[b]
 \includegraphics[width= 8cm]{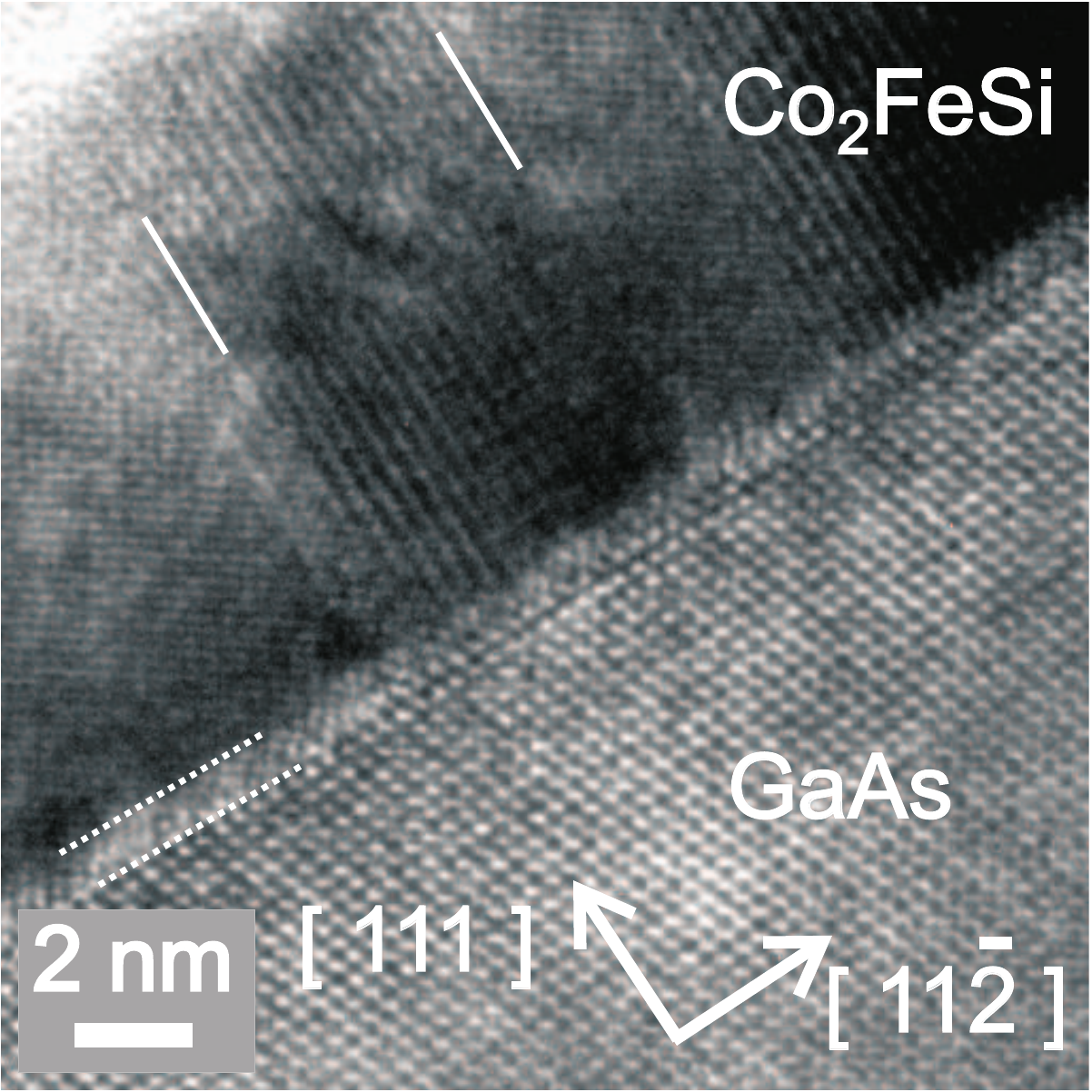}

\caption{Cross-section HRTEM micrograph  along the
GaAs~{[}0$1\bar{1}${]} zone axis of a sample grown at
$T_{S}$~=~275$\thinspace^{\circ}$C. The FM/SC interface is
slightly modified (see region between dashed lines) compared to
those of samples grown at lower T$_{S}$. Inhomogeneities of the
Co$_{2}$FeSi film are visible (marked by full lines).}

\label{fig:figure6}
\end{figure}

%
\begin{figure}[b]
 \includegraphics[width= 8cm]{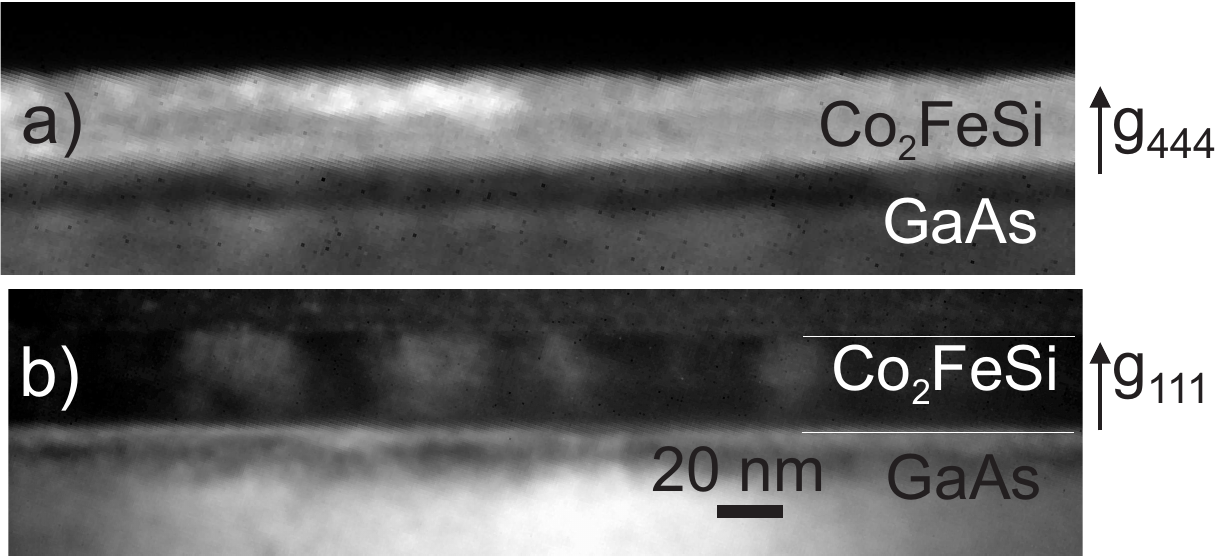}

\caption{Dark-field cross-section TEM micrographs of the sample
grown at $T_{S}$~=~275$\thinspace^{\circ}$C. a) The fundamental
Co$_{2}$FeSi 444 reflection diffracts quite intensively and
homogeneously. b) The image of the  Co$_{2}$FeSi 111 superlattice
reflection of the same region is less intense and less
homogeneous. On the GaAs side of the FM/SC interface both
reflections reveal contrasts due to lattice strain connected to
diffusion into the GaAs substrate. }

\label{fig:figure7}
\end{figure}

%
\begin{figure}[b]
 \includegraphics[width= 8cm]{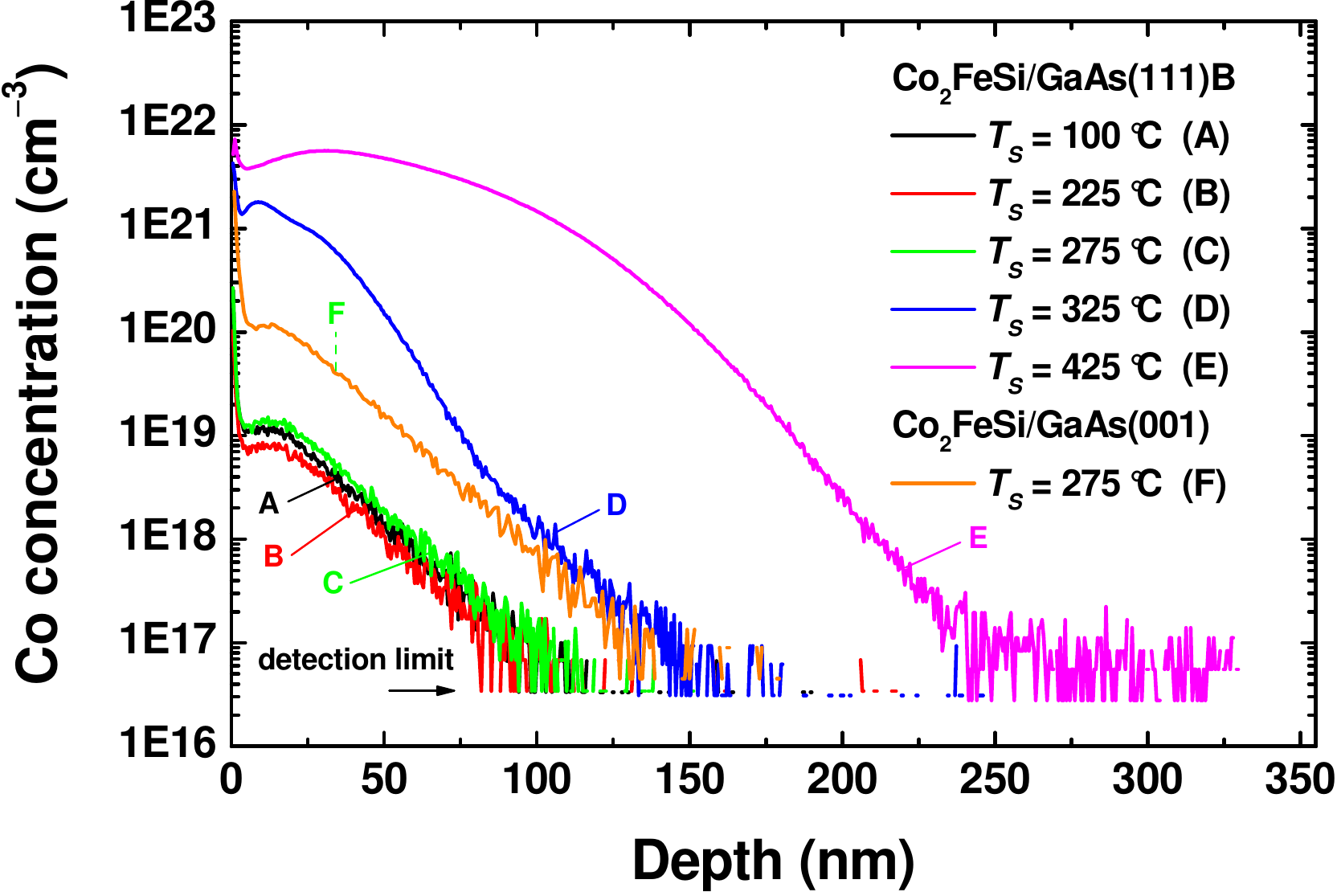}

\caption{SIMS depth profiles for Co in-diffusion at different
growth temperatures of the Co$_{2}$FeSi layers. }

\label{fig:figure8}
\end{figure}

%
\begin{figure}[b]
 \includegraphics[width= 8cm]{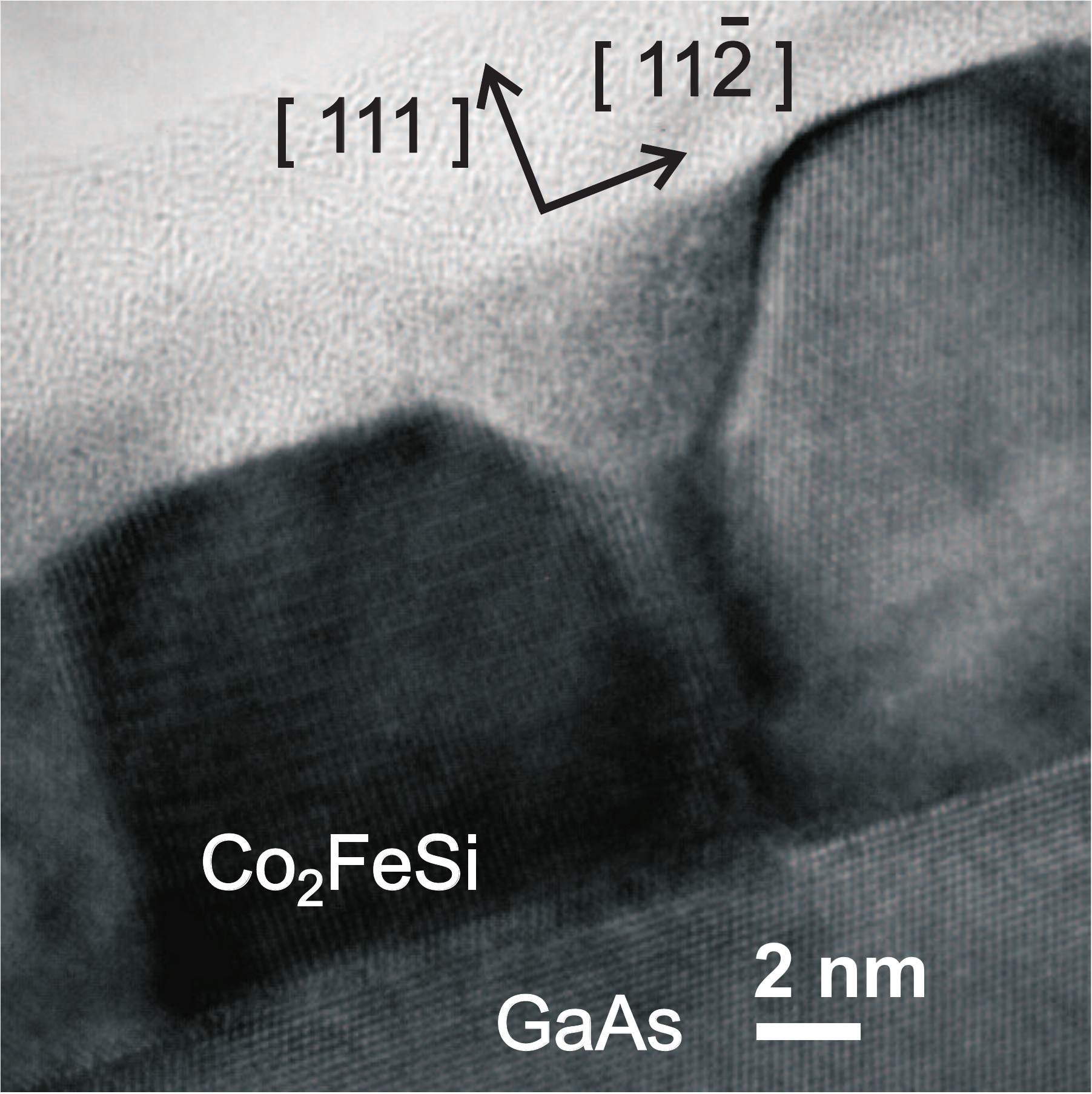}

\caption{Cross-section HRTEM micrograph along the
GaAs~{[}0$1\bar{1}${]} zone axis of a sample grown at
$T_{S}$~=~425$\thinspace^{\circ}$C. Different grains can be
distinguished clearly and for one grain a modulated contrast
pattern is observed. Obviously the grains were formed after
coalescence of islands. The shape of the former islands is visible
from the shape of the Co$_{2}$FeSi surface. }

\label{fig:figure9}
\end{figure}


\begin{figure}[b]
 \includegraphics[width= 8cm]{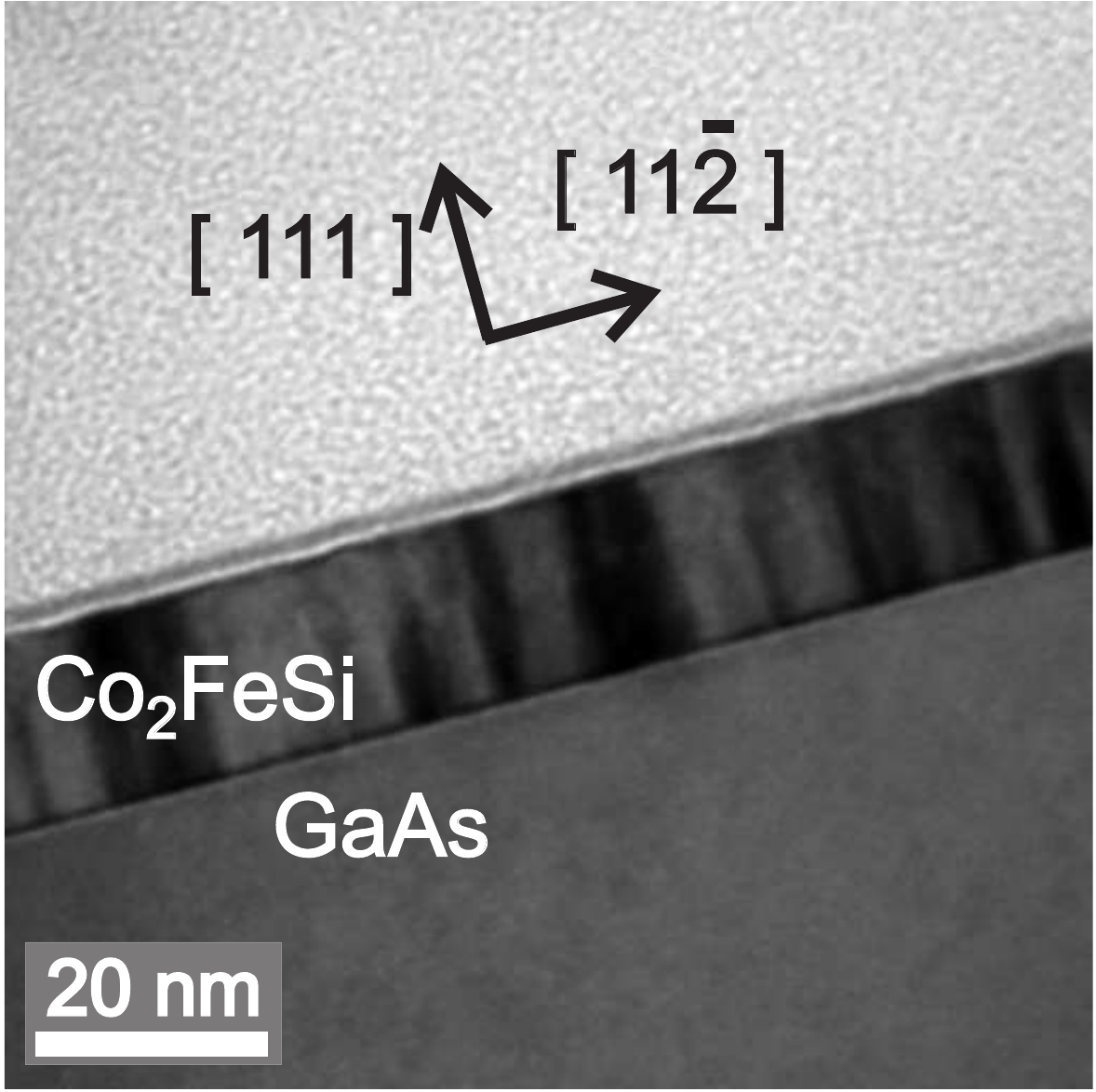}

\caption{Cross-section multi-beam TEM micrograph at lower
magnification along the GaAs~{[}0$1\bar{1}${]} zone axis of a
sample grown at $T_{S}$~=~100$\thinspace^{\circ}$C. Different
grains can be distinguished due to changing contrast in the
micrograph. The GaAs substrate diffracts homogeneously. }

\label{fig:figure10}
\end{figure}


\begin{figure}[b]
 \includegraphics[width= 8cm]{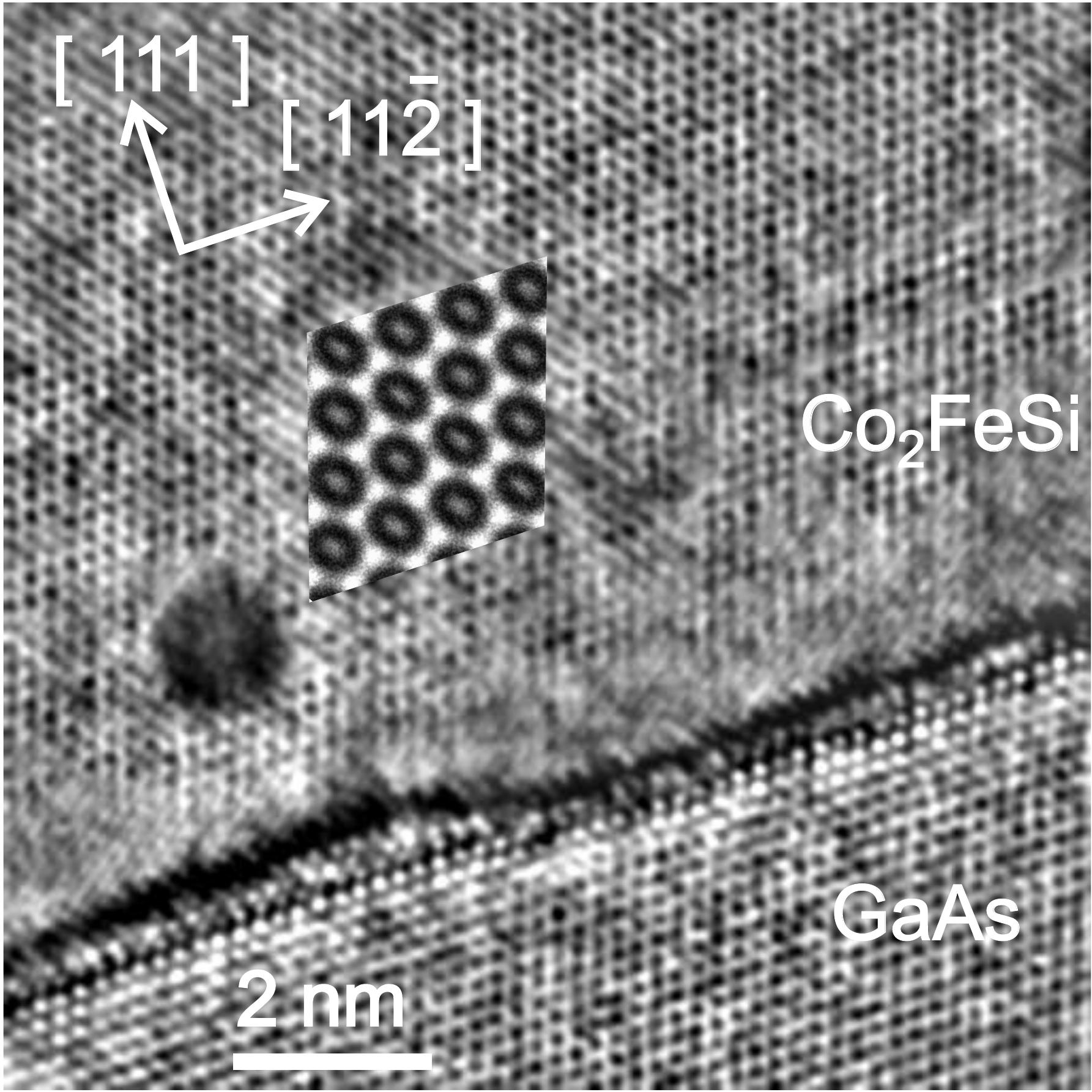}

\caption{Cross-section HRTEM micrograph (Fourier filtered) along
the GaAs~{[}0$1\bar{1}${]} zone axis of a sample grown at
$T_{S}$~=~425$\thinspace^{\circ}$C. Here the dominant $L2_{1}$
phase of Co$_{2}$FeSi is observed more far from the interface
(inset: simulation), whereas the $B2$ phase (fringed contrast) is
seen near the interface. }

\label{fig:figure11}
\end{figure}

\section{Experimental}

The structures were grown by solid-source MBE. The growth
procedure on GaAs(001) is described in detail in Refs.
\cite{hashimoto05,hashimoto05JAP}. In our case GaAs(111)B samples
with a 50-nm-thick GaAs buffer layer and an As-stabilized,
($2$$\times$$2$)-reconstructed surface were grown first on a
GaAs(111)B substrate and then transferred to an As-free growth
chamber, in which stoichiometric Co$_{2}$FeSi films were grown on
top for different substrate temperatures $T_{S}$ ranging between
$T_{S}$~=~100$\thinspace^{\circ}$C and 425$\thinspace^{\circ}$C.
Co, Fe, and Si are co-deposited from high-temperature effusion
cells. The evaporation rates are controlled by the cell
temperatures and are in accordance with the optimized fluxes for
the growth of stoichiometric Co$_{2}$FeSi films on GaAs(001)
substrates \cite{hashimoto05}. The growth rate was approximately
one monolayer in three minutes. The base pressure was about
$1\times10^{-10}$~Torr. The growth was monitored by
\textit{in-situ} reflection high energy electron diffraction
(RHEED). The samples were investigated by dark-field and
high-resolution (HR) TEM. For that purpose cross-sectional TEM
specimens were prepared by mechanical lapping and polishing,
followed by argon ion milling according to standard techniques.
TEM images were acquired with a JEOL 3010 microscope operating at
300 kV. In addition, X-ray measurements of diffraction and
reflectivity were performed in a Panalytical X`Pert diffractometer
equipped with a Ge 220 Hybrid monochromator using $CuK{\alpha}_1$
radiation. Some of the samples were characterized by
\textit{ex-situ} AFM almost immediately after the growth. A
Digital Instrument Nanoscope was utilized for this purpose. Depth
profiles of the film constituents inside the GaAs buffer layer
were determined by secondary ion mass spectometry (SIMS). The
Co$_{2}$FeSi layer was removed by selective wet chemical etching
prior the SIMS profiling. The SIMS measurements were performed
with a CAMECA IMS4F system.

\section{Results}

Fully ordered Co$_{2}$FeSi crystallizes into the face centered
$L2_{1}$ structure. This structure can be viewed as an fcc lattice
with a basis consisting of four atoms A, B, C, and D with
coordinates A(0, 0, 0), B(0.25, 0.25, 0.25), C(0.5, 0.5, 0.5), and
D(0.75, 0.75, 0.75). In the ordered Co$_{2}$FeSi crystal, Co atoms
occupy two sublattices A, C, and Fe is on sublattice B, while Si
atoms fill the sublattice D. Fe and Co atoms cannot be
distinguished by TEM, because they have very similar scattering
factors ($f_{\mathrm{Fe}}\approx f_{\mathrm{Co}}$). In this
approximation the degree of disorder can be described by two order
parameters $\alpha $ and $\beta $, which are fractions of Si atoms
occupying the Fe(B) and Co(A,C) sites, respectively. Three types
of reflections are expected for such a lattice: Fundamental
reflections (not sensitive to disorder) and the two kinds of
superlattice reflections arising due to long-range order.
Figure~\ref{fig:figure1} shows a selected area electron
diffraction pattern of the Co$_{2}$FeSi film along the
{[}0$1\bar{1}${]} direction. We see a rectangular grid of the
fundamental reflections (indexed, see red line) and a less intense
hexagonal grid (see black line) of superlattice reflections in the
vicinity.

Fundamental reflections are not influenced by disorder. They are
given by the rule $H+K+L=4n$, where $n$ is an integer. One example
for a fundamental reflection is 444. The structure amplitude is
given by
\begin{equation}
F_{444}=4(f_{\mathrm{Si}}+f_{\mathrm{Fe}}+2f_{\mathrm{Co}}),
\label{eq1}
\end{equation}
where $f_{\mathrm{Si}}$,  $f_{\mathrm{Fe}}$, $f_{\mathrm{Co}}$ are
atomic scattering factors of the respective elements. There are
two distinct types of superlattice reflections \cite
{niculescu76}. Reflections with odd $H,K,L$  like the 111
reflection are sensitive to both types of disorder (in the Fe(B)
and Co(A,C) sublattices with the corresponding order parameters
$\alpha $ and $\beta $), the structure amplitude being
\begin{equation}
F_{111}=4i(1-2\alpha -\beta )(f_{\mathrm{Si}}-f_{\mathrm{Fe}}).
\label{eq2}
\end{equation}
Reflections satisfying the condition $H+K+L=4n-2$ are sensitive
only to disorder in the Co(A,C) sublattice, an example is 222 with
the structure amplitude
\begin{equation}
F_{222}=-4(1-2\beta )(f_{\mathrm{Si}}-f_{\mathrm{Co}}).
\label{eq3}
\end{equation}
Compared to the fully ordered $L2_{1}$ phase, the $B2$ phase is
obtained by a complete mixing between the Fe and the Si atoms in
the Co$_{2}$FeSi lattice. The Co(A,C) sublattice remains fully
ordered in this case and the 222 reflection has still a high
intensity, see equation (\ref{eq3}).

Figure~\ref{fig:figure2} shows high-resolution x-ray diffraction
curves near the 222 diffraction peak of some of the samples
investigated. Besides the narrow quasi-forbidden GaAs reflection,
the broader Co$_{2}$FeSi maximum is pronounced together with
thickness fringes. Those fringes indicate a smooth FM/SC interface
and and a high quality surface of the film, similar like the
results of x-ray reflectivity measurements. The observation of an
intense 222 reflection indicates that the Co(A,C) sublattice
remains fully ordered and $\beta \approx 0$, see equation
(\ref{eq3}). That means the Co$_{2}$FeSi film shows at least a
$B2$ ordering, i.e. the Si atoms can be mixed only with the Fe
atoms ($\alpha > 0$) and not with the Co atoms ($\beta \approx
0$). Modifications of the diffraction curves are found for
$T_{S}=325~^{\circ}$C where the thickness fringes are less
pronounced and $T_{S}=425~^{\circ}$C where a shift of the peak is
found and the thickness fringes vanish.

The morphology of the Co$_{2}$FeSi surface after growth at
T$_{S}$~=~100$\thinspace^{\circ}$C imaged by atomic force
microscopy is excellent, see Fig.~\ref{fig:figure3} (a). For
higher growth temperatures a slight deterioration of the surface
is found, although the root mean squared (RMS) roughness is
limited to values below 0.5~nm for growth temperatures up to
$T_{S}$~=~275$\thinspace^{\circ}$C, Fig.~\ref{fig:figure3} (b).

Figure~\ref{fig:figure4} shows a cross-section HRTEM micrograph
(Fourier filtered) along the GaAs {[}2$\bar{1}\bar{1}${]} zone
axis of a sample grown at $T_{S}$~=~100$\thinspace^{\circ}$C and a
corresponding selected area diffraction pattern illustrating the
cube-on-cube orientational relationship between Co$_{2}$FeSi and
GaAs. All the diffraction maxima of Co$_{2}$FeSi and GaAs overlap
fully due to the vanishing misfit between stoichiometric
Co$_{2}$FeSi and GaAs. For all samples investigated, the
orientational relationship $(111)$GaAs $\parallel$
$(111)$Co$_{2}$FeSi and $\langle110\rangle$GaAs $\parallel$
$\langle110\rangle$Co$_{2}$FeSi was established. The interface
between the Co$_{2}$FeSi film and the GaAs buffer layer is smooth
and distributed only over a few monolayers.

Figure~\ref{fig:figure5} is a multi-beam cross-section TEM
micrograph at lower magnification along the GaAs {[}0$1\bar{1}${]}
zone axis of a Co$_{2}$FeSi film  grown on GaAs at
$T_{S}$~=~325$\thinspace^{\circ}$C. A large reacted region (lower
edge marked by dotted line) is visible below the
Co$_{2}$FeSi/GaAs(111) interface. Similar elongated precipitates
produced by interfacial reactions were detected by cross-sectional
TEM for all samples grown at $T_{S}\geq325~^{\circ}$C.
Additionally, we found peaks in x-ray diffraction curves caused by
the interface reaction products. The precipitates in the samples
grown at $T_{S}$~=~325$\thinspace^{\circ}$C are always elongated
along the interface and are found only in the immediate vicinity
of the interface. They are detected mainly in the GaAs buffer
layer, and their predominant boundary is (111). A typical lateral
size of these precipitates is $\approx$100~nm, whereas the extent
of the precipitates perpendicular to the interface is about
10--20~nm, i.e., it may sometimes even reach the film thickness
like shown in Fig.~\ref{fig:figure5}. The distance between the
precipitates is of the order of micrometers. The remaining part of
the FM/SC interface between the precipitates is almost perfect.
The precipitates in the samples grown at
$T_{S}$~=~425$\thinspace^{\circ}$C are even larger and have
irregular shapes.

The samples grown at $T_{S}=225~^{\circ}$C  and at
$T_{S}=275~^{\circ}$C were still free of interface reactions. For
$T_{S}=275~^{\circ}$C  we could observe some additional phenomena:
In the HRTEM micrograph (Fig.~\ref{fig:figure6}) a slight
modification of the interface (see e.g. the region between the
dashed lines) is visible together with the typical inhomogeneities
(marked by full lines) of the Co$_{2}$FeSi films. The origin of
these inhomogeneities will be explained later.
Figure~\ref{fig:figure7} shows a comparison of two dark-field
micrographs: taken (a) with the 444 fundamental reflection (not
sensitive to disorder of the Co$_{2}$FeSi), and (b) with the 111
superlattice reflection (sensitive to disorder).

Let us first consider the GaAs buffer layer: On the GaAs side of
the interface we see in (a) and (b) an inhomogeneous stripe caused
by strain connected to diffusion of film atoms into the GaAs
buffer layer. This diffusion was also detected by SIMS depth
profiling, as shown in Fig.~\ref{fig:figure8} for the example of
Co in-diffusion. The maximum Co concentration in the uppermost
part of the GaAs is almost constant up to a temperature of
$T_{S}$~=~275$\thinspace^{\circ}$C and then increases drastically
for the higher growth temperatures. Similar results are obtained
for the Fe and Si depth profiles (not shown here). For
$T_{S}$~$\leq$~325$\thinspace^{\circ}$C, a maximum in the Co
concentration is found at approximately 20 nm below the interface,
which is in good agreement with the observation in
Fig.~\ref{fig:figure7}. A Co concentration depth profile below a
Co$_{2}$FeSi layer grown on GaAs(001) at
$T_{S}$~=~275$\thinspace^{\circ}$C is also included in
Fig.~\ref{fig:figure8}. However, the diffusion of Co into the GaAs
in this case is strongly enhanced by almost one order of magnitude
compared to the films grown on GaAs(111)B at the same T$_{S}$
verifying the reduced diffusion activity perpendicular to the
(111) plane.

Now let us consider the Co$_{2}$FeSi film: The 444 reflection
shown in Fig.~\ref{fig:figure7} (a) is a fundamental reflection
for the Co$_{2}$FeSi lattice, i.e. it is not sensitive to
disorder. The disorder-sensitive superlattice 111 reflection is
imaged in Fig.~\ref{fig:figure7} (b)
\cite{niculescu76,niculescu79}. Under the condition, that the TEM
micrograph with g~=~444 is homogeneous the micrograph taken with
g~=~111 reveals the spatial distribution of long-range order. It
shows brighter areas, that are due to well ordered regions in the
film. These ordered regions are not distributed uniformly.
Moreover a darker stripe is found near the interface. The image of
the Co$_{2}$FeSi 111 reflection is less homogeneous and less
intense than the image of the fundamental Co$_{2}$FeSi 444
reflection shown in Fig.~\ref{fig:figure7}(a).  The striking
difference in intensity of both reflections (the GaAs intensity
should be taken into account as a reference for a comparison) can
be explained by their different structure factors, see equations
(\ref{eq1}) and (\ref{eq2}). We estimated from x-ray diffraction
results that approximately $25\%$ of the Si atoms left their
sublattice (sample grown at $T_{S}=275~^{\circ}$C). For such a
disorder the Co$_{2}$FeSi 444 reflection is expected to be more
than 50 times more intense than the Co$_{2}$FeSi 111 reflection, a
value, which seems to be in a reasonable agreement with
Fig.~\ref{fig:figure7}. The image of the Co$_{2}$FeSi 222
reflection is more homogeneous than the 111 reflection. This is
another hint, that the inhomogeneities are caused mainly by mixing
of the Fe and the Si atoms in  the Co$_{2}$FeSi.

In the micrograph shown in Fig.~\ref{fig:figure9}, we observe well
distinguished grains originating from island growth of
Co$_{2}$FeSi on GaAs(111)B and the coalescence of those islands.
Grains of different orientation produce different interference
patterns (phase contrast) in the HRTEM micrograph. Additional tilt
or twist of the grains modify these patterns. In the micrograph,
one of the grains seems darker and exhibits a modulated phase
contrast, which is probably caused by a slight tilt or twist of
that grain with respect to its neighbors. In other regions, these
modulation stripes are found to be perpendicular to the FM/SC
interface pointing to a lateral strain in the grains. So the
inhomogeneity in the HRTEM micrograph shown in
Fig.~\ref{fig:figure6} resembles the grain structure of the film,
which is visible in Fig.~\ref{fig:figure10} on a larger scale.
While the GaAs substrate diffracts homogeneously, we see an
inhomogeneous contrast in the micrograph of the film, probably
caused by residual strain in the different grains. Such a strain
could be caused by a local change in stoichiometry, which is
always accompanied by a change in the lattice parameter
\cite{hashimoto05}. A change in local stoichiometry may be due to
a lateral segregation or stress relaxation near the step edges of
the GaAs surface.

Figure~\ref{fig:figure11} shows again a HRTEM micrograph of a
sample grown at higher T$_{S}$~=~425$\thinspace^{\circ}$C. Here we
see an area with clearly dominant well ordered $L2_{1}$ phase in
the Co$_{2}$FeSi film (honeycomb-like contrast, inset: simulation,
performed with the program Electron Microscopy Image Simulation,
EMS On Line, http://cecm.insa-lyon.fr/CIOL/). Only the region very
near to the FM/SC interface consists of the $B2$ phase (fringe
contrast). A comparison of Figures~\ref{fig:figure11} and
~\ref{fig:figure6} demonstrates that the micrograph of the sample
grown at the lower T$_{S}$~=~275$\thinspace^{\circ}$C exhibits
larger areas of fringed contrast patterns with additional
inhomogeneities, i.e. the long-range order of the Co$_{2}$FeSi
lattice is improved for the samples grown at higher $T_{S}$.

\section{Discussion}

At $T_{S}\geq325~^{\circ}$C  a small amount of plate-like
particles are found beneath the FM/SC interface. These particles
are obstacles for spin injection. However, the distance between
these precipitates is about $1~\mu$m or more. The remaining part
of the FM/SC interface is almost structurally perfect and may
still contribute to an effective spin injection into the
semiconductor device. This has been recently demonstrated for
Co$_{2}$FeSi/(Al,Ga)As spin LEDs grown on GaAs(001) substrates,
where a higher spin injection efficiency was obtained inspite of
the presence of precipitates due to the high T$_{S}$ of the spin
injector layer \cite{ramsteiner08}. The morphology of the reacted
regions near the (111) interface is clearly different from the
appearance of the precipitates found in samples grown on GaAs(001)
\cite{hashimoto07}. In our case the precipitates are larger and
more flattened along the (111) interface. Therefore (111)
boundaries seem to act as barriers for diffusion-driven
precipitate formation.

Our samples grown at $T_{S}=275~^{\circ}$C were still free of
interface reactions, although a modification of the interface was
found, see Fig.~\ref{fig:figure6}. MBE growth of Co$_{2}$FeSi on
GaAs(001) resulted in an interface reaction at the lower
$T_{S}=250~^{\circ}$C, while a modification of the interface took
already place at $T_{S}=200~^{\circ}$C , cf.
\cite{hashimoto05,hashimoto07}. So, the Co$_{2}$FeSi/GaAs(111)B
hybrid structures are stable at temperatures 75 K higher than
Co$_{2}$FeSi/GaAs(001) structures in agreement with our
expectation mentioned in the introduction.

Despite a vanishing misfit between the Co$_{2}$FeSi film and the
GaAs buffer layer, the observed formation of grains in the film is
evidence for three-dimensional island growth during the first
stage of heteroepitaxy of Co$_{2}$FeSi on GaAs(111)B before a
continuous film is formed. Such island growth is caused by the
higher surface tension of the metal compared to the semiconductor
substrate. The surface tension of Co$_{2}$FeSi is similar to that
of Fe$_{3}$Si, for which the Volmer-Weber growth mode was observed
on GaAs(001) earlier using \textit{in-situ} x-ray diffraction
\cite{kag09,Jenichen09}. Indeed the difference of the surface
tension $\Delta\gamma_1$ of Fe$_{3}$Si and Co$_{2}$FeSi should be
small compared to the difference $\Delta\gamma_2$ of GaAs and
Co$_{2}$FeSi  ($\Delta\gamma_1\ll\Delta\gamma_2$) leading us to
the expectation  of an island growth mode of Co$_{2}$FeSi on
GaAs(111)B. However, in-situ RHEED measurements did not detect a
roughening of the surface indicating relatively large and very
flat islands. Our growth rate one monolayer in three minutes is
very near to a simulated kinetic optimum of Volmer-Weber growth
with a resulting RMS surface roughness of less than one monolayer
during all stages of the growth, see kinetic Monte Carlo
simulations in
 \cite{kag09}.

We show by HRTEM that the long-range order of the Co$_{2}$FeSi
lattice improves with increasing $T_{S}$. However, near the FM/SC
interface we often find some disordering (B2 phase) especially at
the higher $T_{S}$. In addition, the lateral homogeneity of
ordering can be imaged by dark-field TEM using superlattice
reflections e.g. 111 or 222, provided, the image of the
fundamental reflection 444 is more homogeneous in the same area.
In our case the spatial distribution of the long-range order in
Fig.~\ref{fig:figure7}(b) is
 similar to the distribution of grains shown in Figs.
~\ref{fig:figure9} and~\ref{fig:figure10}. The origin of these
grainy inhomogeneities in the long-range order probably lies in a
segregation at step edges or a local strain relaxation of the the
Co$_{2}$FeSi film (without formation of misfit dislocations) via
non-stoichiometry (with a formation and movement of point defects)
leading to a local change in the lattice. In one way or the other
neighboring grains would have a slightly different lattice
parameter. No extended defects are needed for these
inhomogeneities. A change in stoichiometry of the the Co$_{2}$FeSi
film could arise as a consequence of the diffusion into the GaAs
buffer layer as well. Obviously inhomogeneity of the stoichiometry
is directly connected to inhomogeneity of long-range ordering.
During the epitaxial growth the islands nucleate independently and
randomly on the GaAs surface, grow and eventually coalesce. In
this way the local growth conditions in each of the grains
influence long-range order in this grain, e.g. due to different
sticking of the Co$_{2}$FeSi film atoms at step edges or simply
due to a slight change in stoichiometry driven by inhomogeneity of
local mismatch.

\section{Summary}

Co$_{2}$FeSi films on GaAs(111)B grow in the Volmer-Weber island
growth mode, contain the ordered $L2_{1}$ phase, and have a stable
interface up to $T_{S}=275~^{\circ}$C, a growth temperature which
is 75~K higher than the temperature guaranteeing a stable
interface during growth on GaAs(001). The distribution of
long-range order in the Co$_{2}$FeSi films can be characterized by
comparison of the dark-field TEM images taken in a homogeneous
fundamental reflection and an inhomogeneous superlattice
reflection. The spatial inhomogeneities in the long-range order
resemble the grainy structure of the Co$_{2}$FeSi film. They are
arising during the growth of the islands and their coalescence via
a local change in stoichiometry, which can lead to a relaxation of
local strain without formation of extended defects.

\section{Acknowledgement}

The authors thank Hans-Peter Schoenherr for his support during the
MBE growth, Doreen Steffen for sample preparation, Astrid Pfeiffer
for help in the laboratory, Esperanza Luna and Vladimir Kaganer
for valuable support and helpful discussion.

\bibliographystyle{apsrev} \bibliographystyle{apsrev}


\begin{thebibliography}{26}
\expandafter\ifx\csname
natexlab\endcsname\relax\def\natexlab#1{#1}\fi
\expandafter\ifx\csname bibnamefont\endcsname\relax
  \def\bibnamefont#1{#1}\fi
\expandafter\ifx\csname bibfnamefont\endcsname\relax
  \def\bibfnamefont#1{#1}\fi
\expandafter\ifx\csname citenamefont\endcsname\relax
  \def\citenamefont#1{#1}\fi
\expandafter\ifx\csname url\endcsname\relax
  \def\url#1{\texttt{#1}}\fi
\expandafter\ifx\csname
urlprefix\endcsname\relax\def\urlprefix{URL }\fi
\providecommand{\bibinfo}[2]{#2}
\providecommand{\eprint}[2][]{\url{#2}}

\bibitem[{\citenamefont{Felser and Hillebrands}(2009)}]{Felser09}
\bibinfo{author}{\bibfnamefont{C.}~\bibnamefont{Felser}} \bibnamefont{and}
  \bibinfo{author}{\bibfnamefont{B.}~\bibnamefont{Hillebrands}},
  \bibinfo{journal}{J. Phys. D: Appl. Phys.} \textbf{\bibinfo{volume}{42}},
  \bibinfo{pages}{080301} (\bibinfo{year}{2009}).

\bibitem[{\citenamefont{Prinz}(1998)}]{prinz98}
\bibinfo{author}{\bibfnamefont{G.~A.} \bibnamefont{Prinz}},
  \bibinfo{journal}{Science} \textbf{\bibinfo{volume}{282}},
  \bibinfo{pages}{1660} (\bibinfo{year}{1998}).

\bibitem[{\citenamefont{Fert}(2008)}]{fert08}
\bibinfo{author}{\bibfnamefont{A.}~\bibnamefont{Fert}}, \bibinfo{journal}{Thin
  Solid Films} \textbf{\bibinfo{volume}{517}}, \bibinfo{pages}{2}
  (\bibinfo{year}{2008}).

\bibitem[{\citenamefont{Zuti\'{c} et~al.}(2004)\citenamefont{Zuti\'{c}, Fabian,
  and Sarma}}]{zutic04}
\bibinfo{author}{\bibfnamefont{I.}~\bibnamefont{Zuti\'{c}}},
  \bibinfo{author}{\bibfnamefont{J.}~\bibnamefont{Fabian}}, \bibnamefont{and}
  \bibinfo{author}{\bibfnamefont{S.~D.} \bibnamefont{Sarma}},
  \bibinfo{journal}{Rev. of Mod. Phys.} \textbf{\bibinfo{volume}{76}},
  \bibinfo{pages}{323} (\bibinfo{year}{2004}).

\bibitem[{\citenamefont{Wurmehl et~al.}(2005)\citenamefont{Wurmehl, Fecher,
  Kandpal, Ksenofontov, Felser, Lin, and Morais}}]{wurmehl05}
\bibinfo{author}{\bibfnamefont{S.}~\bibnamefont{Wurmehl}},
  \bibinfo{author}{\bibfnamefont{G.~H.} \bibnamefont{Fecher}},
  \bibinfo{author}{\bibfnamefont{H.~C.} \bibnamefont{Kandpal}},
  \bibinfo{author}{\bibfnamefont{V.}~\bibnamefont{Ksenofontov}},
  \bibinfo{author}{\bibfnamefont{C.}~\bibnamefont{Felser}},
  \bibinfo{author}{\bibfnamefont{H.-J.} \bibnamefont{Lin}}, \bibnamefont{and}
  \bibinfo{author}{\bibfnamefont{J.}~\bibnamefont{Morais}},
  \bibinfo{journal}{Phys. Rev. B} \textbf{\bibinfo{volume}{72}},
  \bibinfo{pages}{184434} (\bibinfo{year}{2005}).

\bibitem[{\citenamefont{Wurmehl et~al.}(2006)\citenamefont{Wurmehl, Fecher,
  Kandpal, Ksenofontov, Felser, and Lin}}]{wurmehl06}
\bibinfo{author}{\bibfnamefont{S.}~\bibnamefont{Wurmehl}},
  \bibinfo{author}{\bibfnamefont{G.~H.} \bibnamefont{Fecher}},
  \bibinfo{author}{\bibfnamefont{H.~C.} \bibnamefont{Kandpal}},
  \bibinfo{author}{\bibfnamefont{V.}~\bibnamefont{Ksenofontov}},
  \bibinfo{author}{\bibfnamefont{C.}~\bibnamefont{Felser}}, \bibnamefont{and}
  \bibinfo{author}{\bibfnamefont{H.-J.} \bibnamefont{Lin}},
  \bibinfo{journal}{Appl. Phys. Lett.} \textbf{\bibinfo{volume}{88}},
  \bibinfo{pages}{032503} (\bibinfo{year}{2006}).

\bibitem[{\citenamefont{Ohno et~al.}(1999)\citenamefont{Ohno, Young, Beschoten,
  Matsukura, Ohno, and Awschalom}}]{Ohno99}
\bibinfo{author}{\bibfnamefont{Y.}~\bibnamefont{Ohno}},
  \bibinfo{author}{\bibfnamefont{D.~K.} \bibnamefont{Young}},
  \bibinfo{author}{\bibfnamefont{B.}~\bibnamefont{Beschoten}},
  \bibinfo{author}{\bibfnamefont{F.}~\bibnamefont{Matsukura}},
  \bibinfo{author}{\bibfnamefont{H.}~\bibnamefont{Ohno}}, \bibnamefont{and}
  \bibinfo{author}{\bibfnamefont{D.~D.} \bibnamefont{Awschalom}},
  \bibinfo{journal}{Nature} \textbf{\bibinfo{volume}{402}},
  \bibinfo{pages}{790} (\bibinfo{year}{1999}).

\bibitem[{\citenamefont{Zhu et~al.}(2001)\citenamefont{Zhu, Ramsteiner,
  Kostial, Wassermeier, Sch\"onherr, and Ploog}}]{zhu01}
\bibinfo{author}{\bibfnamefont{H.~J.} \bibnamefont{Zhu}},
  \bibinfo{author}{\bibfnamefont{M.}~\bibnamefont{Ramsteiner}},
  \bibinfo{author}{\bibfnamefont{H.}~\bibnamefont{Kostial}},
  \bibinfo{author}{\bibfnamefont{M.}~\bibnamefont{Wassermeier}},
  \bibinfo{author}{\bibfnamefont{H.-P.} \bibnamefont{Sch\"onherr}},
  \bibnamefont{and} \bibinfo{author}{\bibfnamefont{K.~H.} \bibnamefont{Ploog}},
  \bibinfo{journal}{Phys. Rev. Lett.} \textbf{\bibinfo{volume}{87}},
  \bibinfo{pages}{016601} (\bibinfo{year}{2001}).

\bibitem[{\citenamefont{Hanbicki et~al.}(2002)\citenamefont{Hanbicki, Jonker,
  Itskos, Kioseoglou, and Petrou}}]{Hanbicki02}
\bibinfo{author}{\bibfnamefont{A.~T.} \bibnamefont{Hanbicki}},
  \bibinfo{author}{\bibfnamefont{B.~T.} \bibnamefont{Jonker}},
  \bibinfo{author}{\bibfnamefont{G.}~\bibnamefont{Itskos}},
  \bibinfo{author}{\bibfnamefont{G.}~\bibnamefont{Kioseoglou}},
  \bibnamefont{and} \bibinfo{author}{\bibfnamefont{A.}~\bibnamefont{Petrou}},
  \bibinfo{journal}{Appl. Phys. Lett.} \textbf{\bibinfo{volume}{80}},
  \bibinfo{pages}{1240} (\bibinfo{year}{2002}).

\bibitem[{\citenamefont{Ramsteiner et~al.}(2008)\citenamefont{Ramsteiner,
  Brandt, Flissikowski, Grahn, Hashimoto, Herfort, and Kostial}}]{ramsteiner08}
\bibinfo{author}{\bibfnamefont{M.}~\bibnamefont{Ramsteiner}},
  \bibinfo{author}{\bibfnamefont{O.}~\bibnamefont{Brandt}},
  \bibinfo{author}{\bibfnamefont{T.}~\bibnamefont{Flissikowski}},
  \bibinfo{author}{\bibfnamefont{H.~T.} \bibnamefont{Grahn}},
  \bibinfo{author}{\bibfnamefont{M.}~\bibnamefont{Hashimoto}},
  \bibinfo{author}{\bibfnamefont{J.}~\bibnamefont{Herfort}}, \bibnamefont{and}
  \bibinfo{author}{\bibfnamefont{H.}~\bibnamefont{Kostial}},
  \bibinfo{journal}{Phys. Rev. B} \textbf{\bibinfo{volume}{78}},
  \bibinfo{pages}{121303} (\bibinfo{year}{2008}).

\bibitem[{\citenamefont{Sugahara and Tanaka}(2004)}]{Sugahara04}
\bibinfo{author}{\bibfnamefont{S.}~\bibnamefont{Sugahara}} \bibnamefont{and}
  \bibinfo{author}{\bibfnamefont{M.}~\bibnamefont{Tanaka}},
  \bibinfo{journal}{Appl. Phys. Lett.} \textbf{\bibinfo{volume}{84}},
  \bibinfo{pages}{2307} (\bibinfo{year}{2004}).

\bibitem[{\citenamefont{Kallmayer et~al.}(2006)\citenamefont{Kallmayer, Elmers,
  Balke, Wurmehl, Emmerling, Fecher, and Felser}}]{Kallmayer06}
\bibinfo{author}{\bibfnamefont{M.}~\bibnamefont{Kallmayer}},
  \bibinfo{author}{\bibfnamefont{H.~J.} \bibnamefont{Elmers}},
  \bibinfo{author}{\bibfnamefont{B.}~\bibnamefont{Balke}},
  \bibinfo{author}{\bibfnamefont{S.}~\bibnamefont{Wurmehl}},
  \bibinfo{author}{\bibfnamefont{F.}~\bibnamefont{Emmerling}},
  \bibinfo{author}{\bibfnamefont{G.~H.} \bibnamefont{Fecher}},
  \bibnamefont{and} \bibinfo{author}{\bibfnamefont{C.}~\bibnamefont{Felser}},
  \bibinfo{journal}{J. Phys. D: Appl. Phys.} \textbf{\bibinfo{volume}{39}},
  \bibinfo{pages}{786} (\bibinfo{year}{2006}).

\bibitem[{\citenamefont{Attema et~al.}(2006)\citenamefont{Attema, de~Wijs, and
  de~Groot}}]{Attema06}
\bibinfo{author}{\bibfnamefont{J.~J.} \bibnamefont{Attema}},
  \bibinfo{author}{\bibfnamefont{G.~A.} \bibnamefont{de~Wijs}},
  \bibnamefont{and} \bibinfo{author}{\bibfnamefont{R.~A.}
  \bibnamefont{de~Groot}}, \bibinfo{journal}{J. Phys. D: Appl. Phys.}
  \textbf{\bibinfo{volume}{39}}, \bibinfo{pages}{793} (\bibinfo{year}{2006}).

\bibitem[{\citenamefont{Hashimoto et~al.}(2007)\citenamefont{Hashimoto,
  Herfort, Trampert, Sch\"onherr, and Ploog}}]{hashimoto07}
\bibinfo{author}{\bibfnamefont{M.}~\bibnamefont{Hashimoto}},
  \bibinfo{author}{\bibfnamefont{J.}~\bibnamefont{Herfort}},
  \bibinfo{author}{\bibfnamefont{A.}~\bibnamefont{Trampert}},
  \bibinfo{author}{\bibfnamefont{H.-P.} \bibnamefont{Sch\"onherr}},
  \bibnamefont{and} \bibinfo{author}{\bibfnamefont{K.~H.} \bibnamefont{Ploog}},
  \bibinfo{journal}{J. Phys. D: Appl. Phys.} \textbf{\bibinfo{volume}{40}},
  \bibinfo{pages}{1631} (\bibinfo{year}{2007}).

\bibitem[{\citenamefont{Niculescu et~al.}(1976)\citenamefont{Niculescu, Raj,
  Budnick, Burch, Hines, and Menotti}}]{niculescu76}
\bibinfo{author}{\bibfnamefont{V.}~\bibnamefont{Niculescu}},
  \bibinfo{author}{\bibfnamefont{K.}~\bibnamefont{Raj}},
  \bibinfo{author}{\bibfnamefont{J.~I.} \bibnamefont{Budnick}},
  \bibinfo{author}{\bibfnamefont{T.~J.} \bibnamefont{Burch}},
  \bibinfo{author}{\bibfnamefont{W.~A.} \bibnamefont{Hines}}, \bibnamefont{and}
  \bibinfo{author}{\bibfnamefont{A.~H.} \bibnamefont{Menotti}},
  \bibinfo{journal}{Phys. Rev. B} \textbf{\bibinfo{volume}{14}},
  \bibinfo{pages}{4160} (\bibinfo{year}{1976}).

\bibitem[{\citenamefont{Niculescu et~al.}(1979)\citenamefont{Niculescu,
  Budnick, Hines, Raj, Pickard, and Skalski}}]{niculescu79}
\bibinfo{author}{\bibfnamefont{V.}~\bibnamefont{Niculescu}},
  \bibinfo{author}{\bibfnamefont{J.~I.} \bibnamefont{Budnick}},
  \bibinfo{author}{\bibfnamefont{W.~A.} \bibnamefont{Hines}},
  \bibinfo{author}{\bibfnamefont{K.}~\bibnamefont{Raj}},
  \bibinfo{author}{\bibfnamefont{S.}~\bibnamefont{Pickard}}, \bibnamefont{and}
  \bibinfo{author}{\bibfnamefont{S.}~\bibnamefont{Skalski}},
  \bibinfo{journal}{Phys. Rev. B} \textbf{\bibinfo{volume}{19}},
  \bibinfo{pages}{452} (\bibinfo{year}{1979}).

\bibitem[{\citenamefont{Jenichen et~al.}(2005)\citenamefont{Jenichen, Kaganer,
  Herfort, Satapathy, Sch\"onherr, Braun, and Ploog}}]{Jenichen05}
\bibinfo{author}{\bibfnamefont{B.}~\bibnamefont{Jenichen}},
  \bibinfo{author}{\bibfnamefont{V.~M.} \bibnamefont{Kaganer}},
  \bibinfo{author}{\bibfnamefont{J.}~\bibnamefont{Herfort}},
  \bibinfo{author}{\bibfnamefont{D.~K.} \bibnamefont{Satapathy}},
  \bibinfo{author}{\bibfnamefont{H.~P.} \bibnamefont{Sch\"onherr}},
  \bibinfo{author}{\bibfnamefont{W.}~\bibnamefont{Braun}}, \bibnamefont{and}
  \bibinfo{author}{\bibfnamefont{K.~H.} \bibnamefont{Ploog}},
  \bibinfo{journal}{Phys. Rev. B} \textbf{\bibinfo{volume}{72}},
  \bibinfo{pages}{075329} (\bibinfo{year}{2005}).

\bibitem[{\citenamefont{Takamura et~al.}(2009)\citenamefont{Takamura, Nakane,
  and Sugahara}}]{takamura09}
\bibinfo{author}{\bibfnamefont{Y.}~\bibnamefont{Takamura}},
  \bibinfo{author}{\bibfnamefont{R.}~\bibnamefont{Nakane}}, \bibnamefont{and}
  \bibinfo{author}{\bibfnamefont{S.}~\bibnamefont{Sugahara}},
  \bibinfo{journal}{J. Appl. Phys.} \textbf{\bibinfo{volume}{105}},
  \bibinfo{pages}{07109} (\bibinfo{year}{2009}).

\bibitem[{\citenamefont{Bauer}(1958)}]{bauer58}
\bibinfo{author}{\bibfnamefont{E.}~\bibnamefont{Bauer}}, \bibinfo{journal}{Z.
  Kristallographie} \textbf{\bibinfo{volume}{110}}, \bibinfo{pages}{372}
  (\bibinfo{year}{1958}).

\bibitem[{\citenamefont{Tsao}(1992)}]{Tsao1993}
\bibinfo{author}{\bibfnamefont{J.~Y.} \bibnamefont{Tsao}},
  \emph{\bibinfo{title}{Materials Fundamentals of Molecular Beam Epitaxy}}
  (\bibinfo{publisher}{Academic Press, Inc.}, \bibinfo{address}{San Diego, CA},
  \bibinfo{year}{1992}).

\bibitem[{\citenamefont{Kaganer et~al.}(2009)\citenamefont{Kaganer, Jenichen,
  Shayduk, Braun, and Riechert}}]{kag09}
\bibinfo{author}{\bibfnamefont{V.~M.} \bibnamefont{Kaganer}},
  \bibinfo{author}{\bibfnamefont{B.}~\bibnamefont{Jenichen}},
  \bibinfo{author}{\bibfnamefont{R.}~\bibnamefont{Shayduk}},
  \bibinfo{author}{\bibfnamefont{W.}~\bibnamefont{Braun}}, \bibnamefont{and}
  \bibinfo{author}{\bibfnamefont{H.}~\bibnamefont{Riechert}},
  \bibinfo{journal}{Phys. Rev. Lett.} \textbf{\bibinfo{volume}{102}},
  \bibinfo{pages}{016103} (\bibinfo{year}{2009}).

\bibitem[{\citenamefont{Jenichen et~al.}(2009)\citenamefont{Jenichen, Kaganer,
  Shayduk, Braun, and Trampert}}]{Jenichen09}
\bibinfo{author}{\bibfnamefont{B.}~\bibnamefont{Jenichen}},
  \bibinfo{author}{\bibfnamefont{V.~M.} \bibnamefont{Kaganer}},
  \bibinfo{author}{\bibfnamefont{R.}~\bibnamefont{Shayduk}},
  \bibinfo{author}{\bibfnamefont{W.}~\bibnamefont{Braun}}, \bibnamefont{and}
  \bibinfo{author}{\bibfnamefont{A.}~\bibnamefont{Trampert}},
  \bibinfo{journal}{Phys. Stat. Sol. A} \textbf{\bibinfo{volume}{206}},
  \bibinfo{pages}{1740} (\bibinfo{year}{2009}).

\bibitem[{\citenamefont{Hashimoto
  et~al.}(2005{\natexlab{a}})\citenamefont{Hashimoto, Herfort, Sch\"onherr, and
  Ploog}}]{hashimoto05}
\bibinfo{author}{\bibfnamefont{M.}~\bibnamefont{Hashimoto}},
  \bibinfo{author}{\bibfnamefont{J.}~\bibnamefont{Herfort}},
  \bibinfo{author}{\bibfnamefont{H.-P.} \bibnamefont{Sch\"onherr}},
  \bibnamefont{and} \bibinfo{author}{\bibfnamefont{K.~H.} \bibnamefont{Ploog}},
  \bibinfo{journal}{Appl. Phys. Lett.} \textbf{\bibinfo{volume}{87}},
  \bibinfo{pages}{102506} (\bibinfo{year}{2005}{\natexlab{a}}).

\bibitem[{\citenamefont{Hashimoto
  et~al.}(2005{\natexlab{b}})\citenamefont{Hashimoto, Herfort, Sch\"onherr, and
  Ploog}}]{hashimoto05JAP}
\bibinfo{author}{\bibfnamefont{M.}~\bibnamefont{Hashimoto}},
  \bibinfo{author}{\bibfnamefont{J.}~\bibnamefont{Herfort}},
  \bibinfo{author}{\bibfnamefont{H.-P.} \bibnamefont{Sch\"onherr}},
  \bibnamefont{and} \bibinfo{author}{\bibfnamefont{K.~H.} \bibnamefont{Ploog}},
  \bibinfo{journal}{J. Appl. Phys.} \textbf{\bibinfo{volume}{98}},
  \bibinfo{pages}{104902} (\bibinfo{year}{2005}{\natexlab{b}}).



\end{thebibliography}

\end{document}